# Signatures of bilayer Wigner crystals in a transition metal dichalcogenide heterostructure

You Zhou[1,2], Jiho Sung[1,2], Elise Brutschea[1], Ilya Esterlis[2], Yao Wang[2,3], Giovanni Scuri[2], Ryan J. Gelly[2], Hoseok Heo[1,2], Takashi Taniguchi[4], Kenji Watanabe[4], Gergely Zaránd[5], Mikhail D. Lukin[2], Philip Kim[2,6], Eugene Demler[2]† & Hongkun Park[1,2]†

[1]Department of Chemistry and Chemical Biology, Harvard University, Cambridge, MA 02138, USA

[2]Department of Physics, Harvard University, Cambridge, MA 02138, USA

[3]Department of Physics and Astronomy, Clemson University, Clemson, SC 29634, USA

[4]National Institute for Materials Science, 1-1 Namiki, Tsukuba 305-0044, Japan

[5]MTA-BME Quantum Dynamics and Correlations Research Group, Institute of Physics, Budapest University of Technology and Economics Budafoki ut 8., H-1111 Budapest, Hungary

[6]John A. Paulson School of Engineering and Applied Sciences, Harvard University, Cambridge, MA 02138, USA

†To whom correspondence should be addressed: Hongkun_Park@harvard.edu and demler@physics.harvard.edu

**A Wigner crystal, a regular electron lattice arising from strong correlation effects[1-6], is one of the earliest predicted collective electronic states. This many-body state exhibits quantum and classical phase transitions[7] and has been proposed as a basis for quantum information processing applications[8, 9]. In semiconductor platforms, two-dimensional Wigner crystals have been observed under magnetic field[10-17] or moiré-based lattice potential[18-21] where the electron kinetic energy is strongly suppressed. Here, we report bilayer Wigner crystal formation without a magnetic or confinement field in atomically thin MoSe$_2$ bilayers separated by hexagonal boron nitride. We observe optical signatures of robust correlated insulating states formed at symmetric (1:1) and asymmetric (4:1 and 7:1) electron doping of the two MoSe$_2$ layers at cryogenic temperatures. We attribute these features to the bilayer Wigner crystals formed from two commensurate triangular electron lattices in each layer, stabilized via inter-layer interaction[22, 23]. These bilayer Wigner crystal phases are remarkably stable and undergo quantum and thermal melting transitions above a critical electron density of up to $6 \times 10^{12}$ cm$^{-2}$ and at temperatures of ~40 K. Our results demonstrate that atomically thin semiconductors provide a promising new platform for realizing strongly correlated electronic states, probing their electronic and magnetic phase transitions, and developing novel applications in quantum electronics and optoelectronics[24-28].**

Atomically thin heterostructures made of graphene and transition metal dichalcogenide (TMD) monolayers can host a variety of correlated electronic states[29-33]. Recent advances in materials growth and heterostructure fabrication have enabled the preparation of high-quality heterostructures with minimal disorder[34-38]. The large effective masses of charge carriers[39, 40] and the weak Coulomb screening in TMDs suppress the Fermi energy and enhance electron interactions, thus facilitating the realization of correlated electronic phases at much higher electron density[18-21, 41] in comparison to other semiconductors such as GaAs[10-14, 42, 43]. Furthermore, owing to their strong excitonic response that is sensitive to the spin and charge states of the material, the electrical properties of TMD heterostructures can be optically detected[36, 37, 44-46]. Indeed, experiments on TMD heterostructures with moiré superlattices have found optical signatures of correlated electron solids[18-21], sometimes called generalized Wigner crystals[4, 47], aided by the underlying moiré potential at particular electron fillings.

In this work, we create a high-quality $MoSe_2$ bilayer heterostructure and optically investigate the formation of bilayer Wigner crystals at zero magnetic field and high electron densities. The heterostructure consists of two $MoSe_2$ monolayers (nominal 0º twist angle) separated by a 1-nm thick hexagonal boron nitride (hBN) layer (Fig. 1a & b, see methods[48]). The carrier density in $MoSe_2$ layers can be independently controlled by the external gates. Figure 1c shows representative photoluminescence (PL) spectra collected from the $MoSe_2$/hBN/$MoSe_2$ region under various gate configurations at 4K. Without external gating, the PL spectrum is dominated by sharp emission from the neutral intra-layer exciton ($X_0$) and a much weaker signal from the charged exciton ($X_T$). This behavior is similar to high-quality intrinsic monolayers[49], indicating that both $MoSe_2$ layers maintain their direct bandgap (Extended Data. Fig. 1). By doping one layer selectively while keeping the other intrinsic, we find that the neutral excitons in the top and bottom layer are nearly degenerate in energy (within 0.5 meV: Fig. 1c). This observation suggests minimal and near-symmetric strain buildup in the two layers that is different from previous studies[32], facilitating the formation of bilayer Wigner crystals.

Due to the near-degeneracy of the exciton energies in the bilayer, we can characterize the gate-dependent charge states of the system by measuring the reflectance of $X_0$, which is sensitive to the free carrier concentration in MoSe$_2$[36, 37, 44-46] (Fig. 1d, see also Extended Data Fig. 1). Along the line of $V_{tg} = -\alpha V_{bg}$ ($\alpha$ is the ratio between the top and bottom hBN thickness, $\alpha = d_{tg}/d_{bg}$), we observe strong $X_0$ reflectance, indicating that both layers are intrinsic (($i, i$) in Fig. 1d). When $V_{tg}$ and $V_{bg}$ are of the same polarity, both MoSe$_2$ layers are doped and the reflectance is minimized. Importantly, the charge map in Fig. 1d demonstrates full control over the carrier densities in each MoSe$_2$ layer.

Such precise control enables the detailed interrogation of the electronic phases in the heterostructure as a function of charge density. We focus on the ($n, n$) region, where both MoSe$_2$ layers are electron-doped, and map out the $X_0$ reflectance in finer voltage steps (Fig. 2a, corresponding to the dashed box in Fig. 1d). Remarkably, we observe several strong $X_0$ reflectance features, denoted as **I**, **II**, **III**, and **IV** in Fig. 2a, which exist at particular $V_{tg}/V_{bg}$ ratios and diminish at higher voltages. These features are also observable in PL measurements for neutral excitons $X_0$ (Fig. 2b) and charged excitons $X_T$ (Extended Data Fig. 1). Because the reflectance and PL intensity of $X_0$ depends sensitively on the presence of free carriers in the system[36, 37, 44-46], these observations indicate that at these specific $V_{tg}/V_{bg}$ ratios, unexpected insulating states are formed in the bilayer even though both layers are doped with electrons.

Next, we examine the existence conditions for these insulating states. From the PL data and a capacitance model (see Methods), we find that the features **I**, **II**, **III**, and **IV** correspond to electron filling ratios between the top and bottom layers of $n_t:n_b = 1.0(\pm0.2):1$, $4.3(\pm0.4):1$, $7.4(\pm0.7):1$, and 1:4.0(±0.4), respectively ($n_t$ and $n_b$ represents the electron density in the top and bottom layer, respectively). In Fig. 2c we plot the integrated $X_0$ PL intensity as a function of $n_t$ for different $n_b$ values. We observe multiple prominent peaks, corresponding to the features in Fig. 2b, whose positions shift with varying $n_b$. This is in stark contrast with the situation for large $n_b$ (*e.g.,* 4.2 × $10^{12}$ cm$^{-2}$ in Fig. 2c) or what is observed in monolayers: in those situations, the PL intensity of $X_0$ decreases monotonically when the doping is increased. We note that the linewidths of the neutral exciton along features **I**, **II**, **III**, and **IV** are much narrower than those at other $n_t$:$n_b$ ratios or from monolayers of the similar electron density (Fig.2d and Extended Data Fig. 1). These observations indicate that along features **I**, **II**, **III**, and **IV** there is less dephasing stemming from exciton-free carrier interactions[46], thus confirming that these states are more insulating[36, 37, 44-46].

To disentangle the density-dependent behaviour of the insulating states from the intrinsic gate dependence of MoSe$_2$ monolayers[36, 37, 44-46] (Extended Data Fig. 2), we introduce a dimensionless parameter $\delta(n_t, n_b)$ that characterizes the inter-layer coupling:

$$\delta(n_t, n_b) \equiv (I_0(n_t, n_b) - I_t(n_t) - I_b(n_b))/I_0(0, 0). \tag{1}$$

Here $I_0(n_t, n_b)$ is the device's total PL intensity from $X_0$, and $I_t(n_t)$ ($I_b(n_b)$) is the PL intensity of $X_0$ from only the top (bottom) layer when its electron density is $n_t$ ($n_b$). In a situation where there is no interaction between the two MoSe$_2$ layers, the total emission should simply be the sum of individual PL from both layers, and $\delta(n_t, n_b)$ should be zero. The value of $\delta(n_t, n_b)$ therefore quantifies the interactions between the two layers.

Figure 3a presents a two-dimensional (2D) map of $\delta(n_t, n_b)$ extracted from our experimental data in Fig. 2 (see SI for details): it clearly shows the enhancement of $\delta$ along the particular filling ratios $n_t$:$n_b$ compared with other values. Figure 3b shows $\delta$ as a function of total carrier density $n = n_t + n_b$, measured along features **I-IV** in Fig. 2, *i.e.* $n_t$:$n_b$ = 1:1, 4:1, 7:1, and 1:4. The values of $\delta$ reach their maximum at around $10^{12}$ cm$^{-2}$ and diminish to zero at higher densities.

Next, we characterize the temperature dependence of these insulating states. Figures 3c and 3d show the 2D maps of $\delta(n_t, n_b)$ at 23 K and 35 K, which display weaker insulating features compared to 4 K (additional temperature points are shown in Extended Data Figs. 3 & 4). From these temperature-dependent studies, we find that the 1:1, 4:1 and 7:1 features exhibit different thermodynamic stabilities and persist up to 40 K, 30 K, and 23 K, respectively. It is important to note that the critical density for the insulating states also decreases with increasing temperatures, as will be discussed below.

We attribute the unexpected insulating states in Figs. 2 and 3 to the formation of bilayer Wigner crystals. As discussed previously, the large effective masses of charge carriers[39, 40] and weak Coulomb screening in TMD heterostructure are conducive for the Wigner crystal formation[41]. Furthermore, the Fermi energy of a bilayer system is reduced compared to a single layer system with a similar total density because the electron density is divided into two layers (Fig. 4a). Finally, previous theoretical studies have predicted that the inter-layer Coulomb interactions stabilize a bilayer Wigner crystal phase by lowering the system's overall potential energy as compared to two uncoupled single-layer Wigner crystals [22].

To understand the formation of a bilayer Wigner crystal at particular filling ratios, we consider a simple physical model where intra-layer electron-electron interactions are much stronger than those between the layers. In this weak coupling regime, the inter-layer interaction can be treated as a perturbation to two triangular Wigner crystals, and the stable commensurate bilayer lattices form only at particular $n_t$:$n_b$ values. Geometrical considerations dictate that the first few $n_t$:$n_b$ to form such commensurate lattices are 1:1, 3:1, 4:1, and 7:1 (Fig. 4b and SI). We note that in the limit of vanishing inter-layer distance, the $n_t$:$n_b$ values should not be important for the stability of Wigner crystals[50, 51].

To describe our experimental data in a more quantitative fashion, we investigate the system's stability as a function of inter-layer coupling, considering only the potential energy[52], and numerically optimize the lattice geometry for various $n_t$:$n_b$ values. From such analysis, we find that for $n_t$:$n_b$ values of 1:1, 3:1, 4:1, 7:1, the bilayer Wigner crystals have lower energies compared to the decoupled ones due to the commensurability of the two triangular lattices (the layer symmetry dictates that the filling ratios of *e.g.*, 4:1 and 1:4 are equivalent). In contrast, for $n_t$:$n_b$ values that lead to incommensurate stacking of two monolayer Wigner crystals (*e.g.*, 2:1 and 5:1), the bilayer lattices have higher energies than the decoupled layers in the weak coupling regime (Extended Data Figs. 5 & 6). The absence of 3:1, 1:3, and 1:7 features in our experiments suggests that the Wigner crystals at those particular ratios may be unstable in the $MoSe_2$/hBN/$MoSe_2$ system for reasons that we have not accounted for, such as non-ideal contact to TMDs at small gate voltages. Additional experimental and theoretical studies are needed to clarify these issues.

We attribute the density and temperature dependence of the insulating states (Figs. 2 and 3) to the quantum and thermal melting of a Wigner crystal. With increasing electron density, the kinetic energy of the electrons increases faster than the potential energy, eventually melting the Wigner crystal. Notably, at 4 K, the Fermi energy of the electrons near the critical density ($E_F$ ~10 meV near $n_t$ (or $n_b$) ~ $3\times10^{12}$ cm$^{-2}$, based on the electron effective mass[53-55]) is much larger than the thermal energy ($k_B T$ ~ 0.3 meV), and thus the system should exhibit a quantum melting behavior.

Figure 4c shows a 2D map of $\delta(n_t, n_b)$ for $n_t:n_b$ = 1:1 as a function of total electron density $n = n_t + n_b$ and temperature $T$ (see SI for the discussions of $n_t:n_b$ = 4:1 and 7:1 cases). Taking $\delta > 0$ as a proxy for the interaction-driven insulating state, the phase boundary between an electron solid and a liquid can be approximated by the $\delta = 0$ line (dashed line in Fig. 4c). Such an experimentally extracted phase boundary indeed resembles the generic theoretical phase diagram of a Wigner crystal[56], which exhibits a dome shape due to the quantum and thermal phase transitions.

To obtain a qualitative understanding of the bilayer Wigner crystal phase diagram, we calculate the phonon spectrum for the classical bilayer Wigner crystal and estimate the melting curve via a modified Lindemann criterion appropriate to two dimensions[28, 57]. Specifically, we assume that the melting occurs when the ratio of root-mean-square fluctuations of the nearest-neighbor electron distance to the lattice constant exceeds a critical value $\gamma$ (see SI for details). Here we focus on the ratio 1:1. Previous studies have found the Lindemann parameter ($\gamma$) for the bilayer crystal is in fact dependent on the inter-layer separation[58] and we therefore treat it is a fitting parameter. Matching to the $T$ = 4 K critical density, we find $\gamma$ ~ 0.56. We contrast this with the monolayer Wigner crystal, where numerical simulations have shown $\gamma$ ~ 0.1-0.3[28, 59]. The large value of $\gamma$ in our system, which is a direct result of the high critical density observed in the experiment, indicates the enhanced stability of the bilayer crystal.

As shown in Fig. 4d, a melting curve calculated with a modified Lindemann criterion tracks well with the experimentally determined $\delta(n_t, n_b)$ map in Fig. 4c. The dashed curve in Fig. 4d terminates at a critical density ~$1.4\times10^{12}$ cm$^{-2}$, at which point theory predicts the staggered triangular lattice structure becomes unstable, giving way to a sequence of structural transitions[7, 57] (see SI and Extended Data Fig. 5). We emphasize that while the fitting procedure described here yields a melting curve compatible with experimental data, further experimental and theoretical work is needed to understand the mechanism for the dramatic enhancement of stability of the bilayer crystal. We also remark that the temperature scale for melting determined by the experimental data in Fig. 4c is most likely an overestimate of the true melting temperature. Instead, the $\delta = 0$ line in Fig. 4c is more precisely understood as a crossover to the insulating regime, which would precede the establishment of long-range order. The robustness of insulating states for specific $n_t$:$n_b$ values suggests that even in the presence of disorder[60], the Wigner crystal domains are still sufficiently large to establish local crystalline order. Combined optical and transport measurements[61-64] should provide additional information on the interplay between the crystalline correlations and disorder.

The observation of bilayer Wigner crystals in atomically thin semiconductor heterostructures without a magnetic or moiré-confinement field opens up exciting avenues for creating, studying, and manipulating the collective quantum phases of electrons[61-66]. With electrically tunable carrier density and coupling strength, these devices should enable new studies of quantum and classical phase transitions in engineered many-body correlated systems[65, 66]. Unlike a single layer Wigner crystal, which has only one stable lattice structure, bilayer Wigner crystals exhibit a rich structural phase diagram depending on their filling ratios and coupling strength (Extended Data Figs. 5 & 6) and possess unique collective modes such as optical phonons[52, 57, 67, 68]. In addition to providing an experimental platform to investigate the phase transitions of exotic electronic phases in the quantum regime, our devices hold promise for applications in quantum computing[8] and simulation[69] as well. The Wigner crystals realized in atomically thin semiconductors can be controlled and probed both optically and electrically, and can be readily integrated with functional substrates to create on-chip devices for quantum simulations[69-71]. One particularly intriguing direction involves the realization and exploration of tunable spin Hamiltonians that could be used for creating exotic spin states[69-72]. In particular, the strong spin-orbit coupling and valley-dependent selection rules in TMD heterostructures enable optical detection and manipulation of the electronic spin states[73], opening up exciting new directions in simulating and manipulating spin dynamics in various lattice geometries[74, 75].

**Acknowledgments**

We acknowledge support from the DoD Vannevar Bush Faculty Fellowship (N00014-16-1-2825 for HP, N00014-18-1-2877 for PK), NSF CUA (PHY-1125846 for HP, ED, and MDL), Samsung Electronics (for HP and PK), NSF (PHY-1506284 for HP and MDL, DGE-1745303 for EB), AFOSR MURI (FA9550-17-1-0002), ARL (W911NF1520067 for HP and MDL), DOE (DE-SC0020115 for HP and MDL), and BME's TKP 2020 Nanotechnology grant (GZ). The device fabrication was carried out at the Harvard Center for Nanoscale Systems. This research used resources of the National Energy Research Scientific Computing Center (NERSC), a .U.S

Department of Energy Office of Science User Facility operated under Contract No. DE-AC02-05CH11231.

**Author contributions**

H.P. and Y.Z. conceived the project. Y.Z., J.S., and E.B. fabricated the samples and designed/performed the experiments. I.E., Y.W., G.Z., and E.D. developed the theoretical model, and Y.Z., J.S., E.B., I.E., and Y.W. analyzed the data. G.S. assisted with optical measurements, R.G. assisted with sample fabrication, and H.H. grew the $MoSe_2$ crystals. T.T. and K.W. provided hexagonal boron nitride samples. Y.Z., J.S., E.B., I.E., Y.W., G.S, E.D. and H.P. wrote the manuscript with extensive input from the other authors. H.P., E.D., P.K., and M.D.L. supervised the project.

**Competing financial interest**

The authors declare no competing financial interests.

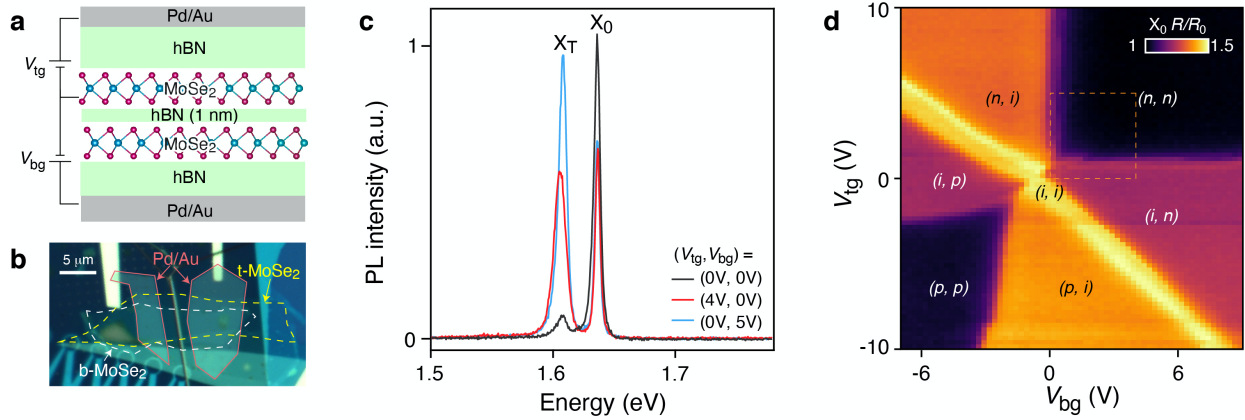

**Figure 1. Device structure and full control of carrier density in the device. a,** A schematic of the heterostructure cross section. The two MoSe$_2$ monolayers are separated by a 1-nm thick layer of hBN. The MoSe$_2$/hBN/MoSe$_2$ structure is then encapsulated in hBN flakes of 15~20 nm thickness. **b,** An optical image of a MoSe$_2$/hBN/MoSe$_2$ device. The top and bottom MoSe$_2$ monolayer regions (t-MoSe$_2$ and b-MoSe$_2$, respectively) are indicated by the yellow and white dashed lines, respectively. The solid red lines show the outline of the optically transparent, thin top Pd/Au gates. **c,** Representative PL spectra of the MoSe$_2$/hBN/MoSe$_2$ device measured at 4 K under different gate configurations. **d,** A 2D map of reflectance contrast from the neutral exciton $X_0$ as a function of top ($V_{tg}$) and bottom ($V_{bg}$) gate voltages at 4 K. We normalize the reflectance $R$ to its value at the highly-doped regime, $R_0$. From the strength of the reflectance, we can extract the charge states of the bilayer system. Here we denote the charge state (t, b) in the order of top and bottom layer. *p, i* and *n* represent hole-doped, neutral and electron-doped, respectively. The dashed box represents the voltage range we focus on in later studies.

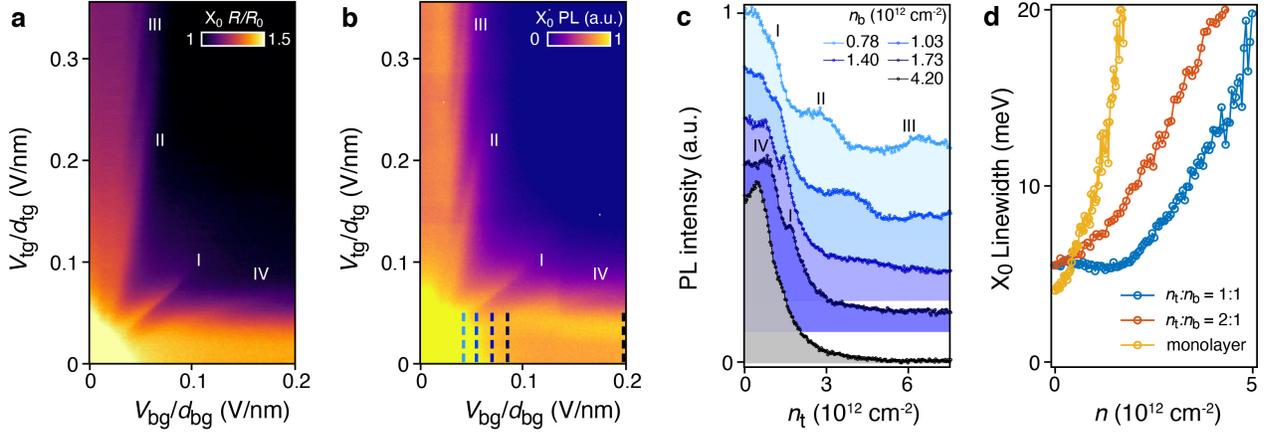

**Figure 2. Voltage-dependent reflectance and PL spectra of a hBN/ MoSe$_2$/hBN/MoSe$_2$/hBN device in the electron doped regime at 4K. a-b,** 2D maps of neutral exciton X$_0$ reflectance contrast $R/R_0$ (**a**) and integrated X$_0$ PL (**b**), as a function of $V_{tg}/d_{tg}$ and $V_{bg}/d_{bg}$. Here we divide the top and bottom gate voltages by the respective hBN thicknesses so that the slopes of the linear features correspond to the ratios of the electrostatically induced carrier densities in each layer, $n_t$:$n_b$. Features denoted as I, II, III, and IV can be clearly observed in these maps. **c,** The integrated X$_0$ PL intensity as a function of $n_t$, while $n_b$ is kept as a constant. The lower bound of the shaded region represents zero PL intensity for each curve. The corresponding bottom gate voltages for these linecuts are indicated by the dashed lines in **b**. **d,** The linewidth of the neutral exciton as a function of total electron density $n$, measured along $n_t$:$n_b$ = 1:1 and 2:1, compared with a monolayer MoSe$_2$.

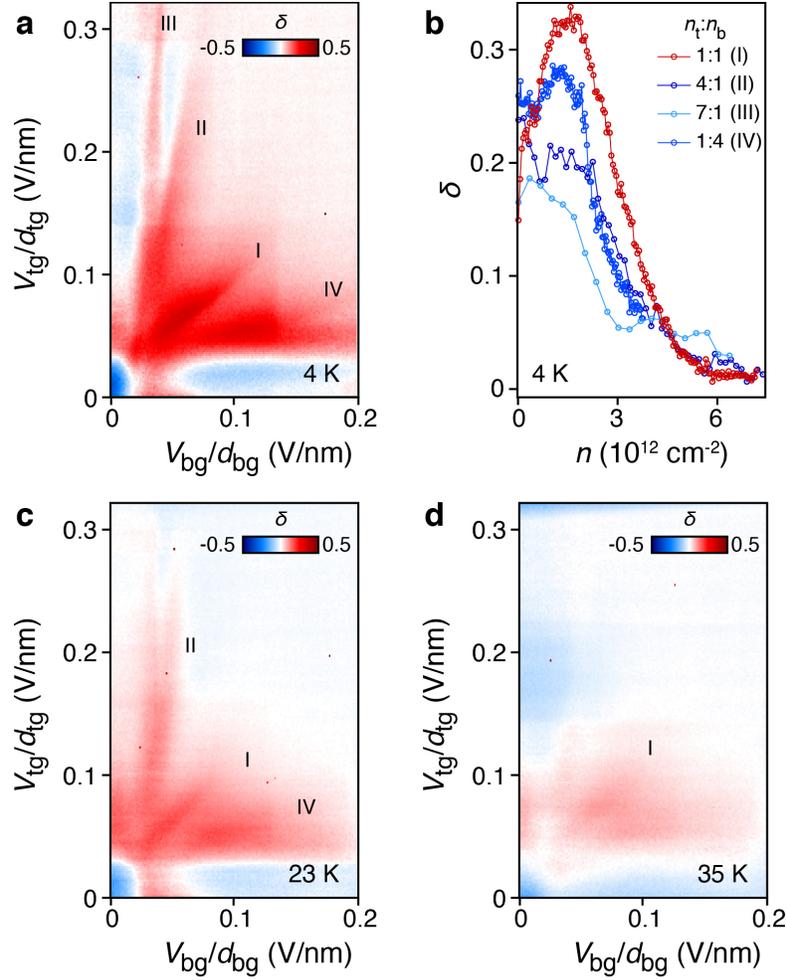

**Figure 3. Density- and temperature-dependence of the interaction-induced insulating states.**
**a,** A 2D map of $\delta(n_t, n_b)$, which characterizies the formation of the interaction-induced insulating states, as a function of $V_{tg}/d_{tg}$ and $V_{bg}/d_{bg}$ at 4K. **b,** $\delta$ as a function of the total electron density in both layers, $n = n_t + n_b$, for various features, *i.e.* $n_t$:$n_b$ = 1:1, 4:1, 7:1, and 1:4 at 4K. **c-d,** 2D maps of $\delta(n_t, n_b)$ as a function of $V_{tg}/d_{tg}$ and $V_{bg}/d_{bg}$ at **c,** 23 K, and **d,** 35K.

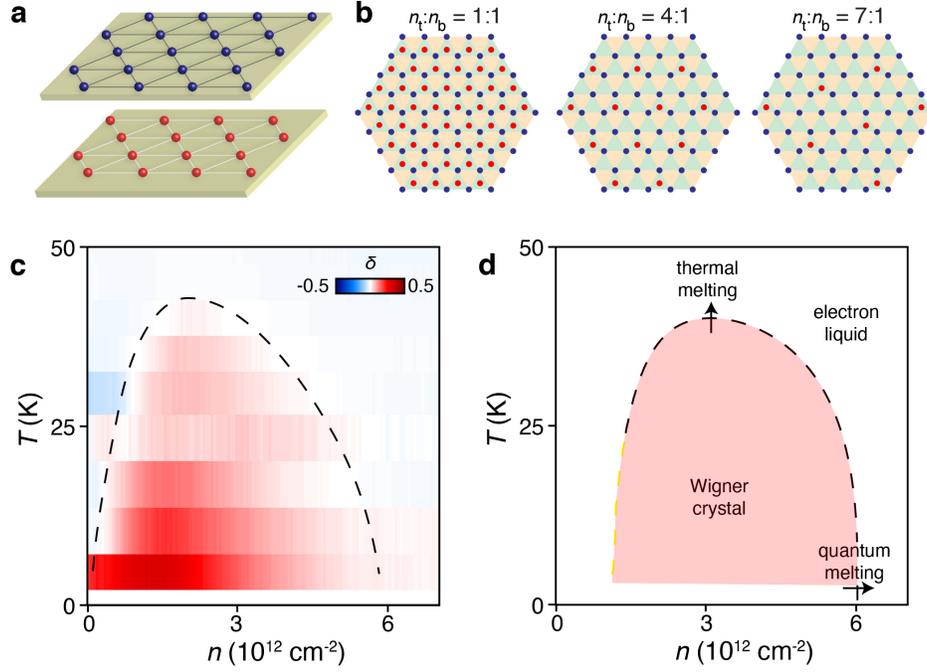

**Figure 4. Bilayer Wigner crystals and their quantum and thermal phase transitions. a,** A schematic of Wigner crystal in bilayer systems. **b,** Schematics of commensurate stacking in bilayer Wigner crystals with triangular lattices for filling ratios $n_t:n_b$ of 1:1, 4:1, and 7:1. **c,** A 2D map of $\delta(n_t, n_b)$ as a function of total carrier density $n$ and temperature $T$ for $n_t:n_b = 1:1$. The $\delta > 0$ region represents where the bilayer Wigner crystal forms (the $\delta = 0$ boundary shown as a dashed line). The center position of each pixel in the $y$ axis represents the temperature. **d,** A schematic phase diagram of bilayer Wigner crystals, showing both quantum and thermal phase transitions. The dashed line represents a calculated melting curve assuming Lindenman parameter $\gamma \sim 0.56$. Below a density of $\sim 1.4 \times 10^{12}$ cm$^{-2}$, theory predicts the staggered triangular lattice structure becomes unstable (green dashed line).

**Methods:**

**Device fabrication**

Monolayer MoSe$_2$ and hBN flakes were exfoliated from bulk crystals onto silicon substrates with a 285-nm silicon oxide layer. Monolayers of MoSe$_2$ were identified under an optical microscope and verified via photoluminescence measurements. The thickness of hBN flakes was measured by an atomic force microscope. Then we fabricated the hBN/MoSe$_2$/hBN/MoSe$_2$/hBN heterostructure using a tear-and-stack technique with a dry transfer method; a monolayer of MoSe$_2$ was torn into two pieces which were then stacked without introducing any rotation between them. This heterostructure was then transferred onto the bottom gate (1 nm Cr and 9 nm Pd/Au alloy), pre-patterned by electron-beam lithography and thermal evaporation. Then top gates (1 nm Cr and 9 nm Pd/Au alloy) were defined with electron-beam lithography and deposited via thermal evaporation. Finally, we made electrical contacts to the MoSe$_2$ and the gates using 5 nm Cr and 90 nm Au deposited via thermal evaporation to connect them to the wire-bonding pads.

**Optical spectroscopy**

Optical measurements were carried out in a home-built confocal microscope with a 4K cryostat from Montana Instruments. The apochromatic objective used has a numerical aperture of 0.75. We excited the sample using a 660-nm wavelength diode laser with various power for the PL measurements. We obtained the reflectance spectra by illuminating the sample with a broadband light source, using a halogen lamp (Thorlabs SLS201L) and a supercontinuum laser from NKT Photonics. The spectra were measured by a spectrometer using either a 300 line/mm or a 1200 line/mm grating and a Princeton Instruments camera (PIXIS 2048). The gate voltages were supplied by two Keithley 2400 sourcemeters.

**Data availability.** The data that support the plots within this paper and other findings of this study are available from the corresponding author upon reasonable request.